\documentstyle[12pt]{article}
\newcommand{\nc}{\newcommand}
\nc{\renc}{\renewcommand}

\nc{\etal}{\mbox{\it et al. }}
\nc{\ie}{{\it i.e.}}
\nc{\eg}{{\it e.g.}}

\renc{\thefootnote}{\arabic{footnote}}
\nc{\capt}[1]{{\bf Figure.} {\small\sl #1}}

\nc{\eqs}[2]{\mbox{Eqs.~(\ref{#1},\,\ref{#2})}}
\nc{\eq}[1]{\mbox{Eq.~(\ref{#1})}}

\nc{\figs}[2]{\mbox{Figs.~(\ref{#1},\,\ref{#2})}}
\nc{\fig}[1]{\mbox{Fig~.(\ref{#1})}}

\nc{\tag}[1]{\label{#1} \marginpar{{\footnotesize #1}}}
\nc{\mtag}[1]{\label{#1} \mbox{\marginpar{{\footnotesize #1}}}}
\renc{\baselinestretch}{1.2}
\newlength{\overeqskip}
\newlength{\undereqskip}
\nc{\be}[1]{\begin{equation} \mbox{$\label{#1}$}}
\nc{\bea}[1]{\begin{eqnarray} \mbox{$\label{#1}$}}
\nc{\Section}[2]{\section{#2}\label{#1}}
\nc{\Bibitem}[1]{\bibitem{#1}}
\nc{\Label}[1]{\label{#1}}

\nc{\eea}{\vspace{\undereqskip}\end{eqnarray}}
\nc{\ee}{\vspace{\undereqskip}\end{equation}}
\nc{\bdm}{\begin{displaymath}}
\nc{\edm}{\end{displaymath}}
\nc{\dpsty}{\displaystyle}
\nc{\bc}{\begin{center}}
\nc{\ec}{\end{center}}
\nc{\ba}{\begin{array}}
\nc{\ea}{\end{array}}
\nc{\bab}{\begin{abstract}}
\nc{\eab}{\end{abstract}}
\nc{\btab}{\begin{tabular}}
\nc{\etab}{\end{tabular}}
\nc{\bit}{\begin{itemize}}
\nc{\eit}{\end{itemize}}
\nc{\ben}{\begin{enumerate}}
\nc{\een}{\end{enumerate}}
\nc{\bfig}{\begin{figure}}
\nc{\efig}{\end{figure}}
\nc{\arreq}{&\!=\!&}
\nc{\arrmi}{&\!-\!&}
\nc{\arrpl}{&\!+\!&}
\nc{\arrap}{&\!\!\!\approx\!\!\!&}
\nc{\non}{\nonumber\\*}
\nc{\align}{\!\!\!\!\!\!\!\!&&}

\def\lsim{\; \raise0.3ex\hbox{$<$\kern-0.75em
      \raise-1.1ex\hbox{$\sim$}}\; }
\def\gsim{\; \raise0.3ex\hbox{$>$\kern-0.75em
      \raise-1.1ex\hbox{$\sim$}}\; }
\nc{\DOT}{\hspace{-0.08in}{\bf .}\hspace{0.1in}}
\nc{\Laada}{\hbox {$\sqcap$ \kern -1em $\sqcup$}}
\nc\loota{{\scriptstyle\sqcap\kern-0.55em\hbox{$\scriptstyle\sqcup$}}}
\nc\Loota{{\sqcap\kern-0.65em\hbox{$\sqcup$}}}
\nc\laada{\Loota}
\nc{\qed}{\hskip 3em \hbox{\BOX} \vskip 2ex}

\nc{\real}{{\rm I \! R}}
\nc{\Z}{{\sf Z \!\!\! Z}}
\nc{\complex}{{\rm C\!\!\! {\sf I}\,\,}}
\def\bigid{\leavevmode\hbox{\small1\kern-3.8pt\normalsize1}}
\def\id{\leavevmode\hbox{\small1\kern-3.3pt\normalsize1}}
\nc{\slask}{\!\!\!/}
\nc{\bis}{{\prime\prime}}
\nc{\pa}{\partial}
\nc{\na}{\nabla}
\nc{\ra}{\rangle}
\nc{\la}{\langle}
\nc{\goto}{\rightarrow}
\nc{\swap}{\leftrightarrow}

\nc{\EE}[1]{ \mbox{$\cdot10^{#1}$} }
\nc{\abs}[1]{\left|#1\right|}
\nc{\at}[2]{\left.#1\right|_{#2}}
\nc{\norm}[1]{\|#1\|}
\nc{\abscut}[2]{\Abs{#1}_{\scriptscriptstyle#2}}
\nc{\vek}[1]{{\rm\bf #1}}
\nc{\integral}[2]{\int\limits_{#1}^{#2}}
\nc{\inv}[1]{\frac{1}{#1}}
\nc{\dd}[2]{{{\partial #1}\over{\partial #2}}}
\nc{\ddd}[2]{{{{\partial}^2 #1}\over{\partial {#2}^2}}}
\nc{\dddd}[3]{{{{\partial}^2 #1}\over
	{\partial #2 \partial #3}}}
\nc{\dder}[2]{{{d #1}\over{d #2}}}
\nc{\ddder}[2]{{{d^2 #1}\over{d {#2}^2}}}
\nc{\dddder}[3]{{d^2 #1}\over
	{d #2 d #3}}
\nc{\dx}[1]{d\,^{#1}x}
\nc{\dy}[1]{d\,^{#1}y}
\nc{\dz}[1]{d\,^{#1}z}
\nc{\dl}[1]{\frac{d\,^{#1}l}{(2\pi)^{#1}}}
\nc{\dk}[1]{\frac{d\,^{#1}k}{(2\pi)^{#1}}}
\nc{\dq}[1]{\frac{d\,^{#1}q}{(2\pi)^{#1}}}

\nc{\cc}{\mbox{$c.c.$ }}
\nc{\hc}{\mbox{$h.c.$ }}
\nc{\cf}{cf.\ }
\nc{\erfc}{{\rm erfc}}
\nc{\Tr}{{\rm Tr\,}}
\nc{\tr}{{\rm tr\,}}
\nc{\pol}{{\rm pol}}
\nc{\sign}{{\rm sign}}
\nc{\bfT}{{\bf T }}

\nc{\cA}{{\cal A}}
\nc{\cB}{{\cal B}}
\nc{\cD}{{\cal D}}
\nc{\cE}{{\cal E}}
\nc{\cG}{{\cal G}}
\nc{\cH}{{\cal H}}
\nc{\cL}{{\cal L}}
\nc{\cO}{{\cal O}}
\nc{\cT}{{\cal T}}
\nc{\cN}{{\cal N}}
\nc{\rvac}[1]{|{\cal O}#1\rangle}
\nc{\lvac}[1]{\langle{\cal O}#1|}
\nc{\rvacb}[1]{|{\cal O}_\beta #1\rangle}
\nc{\lvacb}[1]{\langle{\cal O}_\beta #1 |}
\nc{\bb}{\bar{\beta}}
\nc{\bt}{\tilde{\beta}}
\nc{\ctH}{\tilde{\cal H}}
\nc{\chH}{\hat{\cal H}}
\nc{\1}{\aa}
\nc{\2}{\"{a}}
\nc{\3}{\"{o}}
\nc{\4}{\AA}
\nc{\5}{\"{A}}
\nc{\6}{\"{O}}
\nc{\al}{\alpha}
\nc{\g}{\gamma}
\nc{\Del}{\Delta}
\nc{\e}{\epsilon}
\nc{\eps}{\epsilon}
\nc{\lam}{\lambda}
\nc{\om}{\omega}
\nc{\Om}{\Omega}
\nc{\ve}{\varepsilon}
\nc{\mn}{{\mu\nu}}
\nc{\k}{\kappa}
\nc{\vp}{\varphi}

\nc{\advp}[3]{{\it  Adv.\ in\ Phys.\ }{{\bf #1} {(#2)} {#3}}}
\nc{\annp}[3]{{\it  Ann.\ Phys.\ (N.Y.)\ }{{\bf #1} {(#2)} {#3}}}
\nc{\apl}[3]{{\it  Appl. Phys. Lett. }{{\bf #1} {(#2)} {#3}}}
\nc{\apj}[3]{{\it  Ap.\ J.\ }{{\bf #1} {(#2)} {#3}}}
\nc{\apjl}[3]{{\it  Ap.\ J.\ Lett.\ }{{\bf #1} {(#2)} {#3}}}
\nc{\app}[3]{{\it Astropart.\ Phys.\ }{{\bf #1} {(#2)} {#3}}}
\nc{\cmp}[3]{{\it  Comm.\ Math.\ Phys.\ }{{ \bf #1} {(#2)} {#3}}}
\nc{\cqg}[3]{{\it  Class.\ Quant.\ Grav.\ }{{\bf #1} {(#2)} {#3}}}
\nc{\epl}[3]{{\it  Europhys.\ Lett.\ }{{\bf #1} {(#2)} {#3}}}
\nc{\ijmp}[3]{{\it Int.\ J.\ Mod.\ Phys.\ }{{\bf #1} {(#2)} {#3}}}
\nc{\ijtp}[3]{{\it Int.\ J.\ Theor.\ Phys.\ }{{\bf #1} {(#2)} {#3}}}
\nc{\jmp}[3]{{\it  J.\ Math.\ Phys.\ }{{ \bf #1} {(#2)} {#3}}}
\nc{\jpa}[3]{{\it  J.\ Phys.\ A\ }{{\bf #1} {(#2)} {#3}}}
\nc{\jpc}[3]{{\it  J.\ Phys.\ C\ }{{\bf #1} {(#2)} {#3}}}
\nc{\jap}[3]{{\it J.\ Appl.\ Phys.\ }{{\bf #1} {(#2)} {#3}}}
\nc{\jpsj}[3]{{\it J.\ Phys.\ Soc.\ Japan\ }{{\bf #1} {(#2)} {#3}}}
\nc{\lmp}[3]{{\it Lett.\ Math.\ Phys.\ }{{\bf #1} {(#2)} {#3}}}
\nc{\mpl}[3]{{\it  Mod.\ Phys.\ Lett.\ }{{\bf #1} {(#2)} {#3}}}
\nc{\ncim}[3]{{\it  Nuov.\ Cim.\ }{{\bf #1} {(#2)} {#3}}}
\nc{\np}[3]{{\it  Nucl.\ Phys.\ }{{\bf #1} {(#2)} {#3}}}
\nc{\pr}[3]{{\it Phys.\ Rev.\ }{{\bf #1} {(#2)} {#3}}}
\nc{\pra}[3]{{\it  Phys.\ Rev.\ A\ }{{\bf #1} {(#2)} {#3}}}
\nc{\prb}[3]{{\it  Phys.\ Rev.\ B\ }{{{\bf #1} {(#2)} {#3}}}}
\nc{\prc}[3]{{\it  Phys.\ Rev.\ C\ }{{\bf #1} {(#2)} {#3}}}
\nc{\prd}[3]{{\it  Phys.\ Rev.\ D\ }{{\bf #1} {(#2)} {#3}}}
\nc{\prl}[3]{{\it Phys\ Rev.\ Lett.\ }{{\bf #1} {(#2)} {#3}}}
\nc{\pl}[3]{{\it  Phys.\ Lett.\ }{{\bf #1} {(#2)} {#3}}}
\nc{\prep}[3]{{\it Phys\. Rep.\ }{{\bf #1} {(#2)} {#3}}}
\nc{\prsl}[3]{{\it Proc.\ R.\ Soc.\ London\ }{{\bf #1} {(#2)} {#3}}}
\nc{\ptp}[3]{{\it  Prog.\ Theor.\ Phys.\ }{{\bf #1} {(#2)} {#3}}}
\nc{\ptps}[3]{{\it  Prog\ Theor.\ Phys.\ suppl.\ }{{\bf #1} {(#2)} {#3}}}
\nc{\physa}[3]{{\it  Physica\ A\ }{{\bf #1} {(#2)} {#3}}}
\nc{\physb}[3]{{\it  Physica\ B\ }{{\bf #1} {(#2)} {#3}}}
\nc{\phys}[3]{{\it Physica\ }{{\bf #1} {(#2)} {#3}}}
\nc{\rmp}[3]{{\it  Rev.\ Mod.\ Phys.\ }{{\bf #1} {(#2)} {#3}}}
\nc{\rpp}[3]{{\it Rep.\ Prog.\ Phys.\ }{{\bf #1} {(#2)} {#3}}}
\nc{\sjnp}[3]{{\it Sov.\ J.\ Nucl.\ Phys.\ }{{\bf #1} {(#2)} {#3}}}
\nc{\spjetp}[3]{{\it Sov.\ Phys.\ JETP\ }{{\bf #1} {(#2)} {#3}}}
\nc{\yf}[3]{{\it Yad.\ Fiz.\ }{{\bf #1} {(#2)} {#3}}}
\nc{\zetp}[3]{{\it Zh.\ Eksp.\ Teor.\ Fiz.\  }{{\bf #1}  {(#2)} {#3}}}
\nc{\zp}[3]{{\it Z.\ Phys.\ }{{\bf #1} {(#2)} {#3}}}
\nc{\ibid}[3]{{\sl ibid.\ }{{\bf #1} {#2} {#3}}}
\nc{\rf}[1]{(\ref{#1})}
\nc{\nn}{\nonumber \\*}
\nc{\bfB}{\bf{B}}
\nc{\bfv}{\bf{v}}
\nc{\bfx}{\bf{x}}
\nc{\bfy}{\bf{y}}
\nc{\vx}{\vec{x}}
\nc{\vy}{\vec{y}}
\nc{\oB}{\overline{B}}
\nc{\oI}{\overline{I}}
\nc{\oR}{\overline{R}}
\nc{\rar}{\rightarrow}
\nc{\ti}{\times}
\nc{\slsh}{\hskip-5pt/}
\nc{\sm}{Standard~Model~}
\nc{\MP}{M_{\rm Pl}}
\nc{\tp}{t_{\rm Pl}}
\nc{\ave}{\bar{E}}

\renc{\min}{p_{\rm min}}
\renc{\max}{p_{\rm max}}
\nc{\pmin}{p_{\rm min}}
\nc{\pmax}{p_{\rm max}}
\nc{\fo}{f_0}
\nc{\foi}{f_{0,i}\,}
\nc{\fop}{f_0^P}
\nc{\fou}{f_0^U}

\nc{\eff}{{\rm eff}}
\nc{\MT}{M_{\rm T}}
\nc{\ML}{M_{\rm L}}
\nc{\kk}{\vek{k}}
\nc{\pp}{{\rm p}}
\nc{\cb}{critical bubble~}
\nc{\cbs}{critical bubbles~}
\nc{\scb}{subcritical bubble~}
\nc{\scbs}{subcritical bubbles~}
\nc{\vv}{\\}
\begin{document}

{\title{\vskip-2truecm{\hfill {{\small  SISSA-AP/95-79\\
                }}\vskip 1truecm}
{\bf Are oscillons present during a first order  electroweak phase
transition? }}

\vspace{3cm}

{\author{
{\sc Antonio Riotto$^{\star}$ }\\
{\sl\small International School for Advanced Studies, SISSA-ISAS} \vv
{\sl\small Strada Costiera 11, I-34014, Miramare, Trieste, Italy}\vv
{\sl\small and }\vv {\sl\small Istituto Nazionale di Fisica Nucleare,}
{\sl\small Sezione di Padova, 35100 Padova, Italy}}}

\maketitle
\vspace{2cm}
\begin{abstract}
\noindent
It has been recently argued that localized, unstable, but extremely
long-lived configurations, called oscillons, could affect the
dynamics of a first order electroweak phase transition in an appreciable
way. Treating the amplitude and the size of subcritical bubbles
as statistical degrees of freedom, we show that thermal fluctuations
are not strong enough to
generate  subcritical configurations able to settle into a
an oscillon long-lived  regime.

\end{abstract}
\vfil
\footnoterule
{\small $\star$riotto@tsmi19.sissa.it. Address after November 95:
Theoretical Astrophysics Group, NASA/Fermilab, Batavia, IL60510, USA.}
\thispagestyle{empty}
\newpage
\setcounter{page}{1}

Nontopological soliton solutions of classical field theories were
introduced a number of years ago by Rosen \cite{rosen} and their
properties have been studied by many authors \cite{lee}.

Unlike magnetic monopoles and cosmic strings,
which arise in theories with nontrivial vacuum topology, nontopological
solitons are rendered stable by the existence of a conserved Noether
charge carried by fields confined to a finite region of space. The
minimum charge of the stable soliton depends upon ratios of coupling
constants and in principle can be very small.

Scenarios for actually producing such objects in the early Universe
have also been discussed \cite{gleiser}. In particular,
this issue has been recently rekindled in ref. \cite{osc} where
localized, time-dependent, spherically-symmetric solutions of nonlinear
scalar field theories, called oscillons, were studied and shown to be,
although unstable, extremely long-lived. Indeed, their lifetimes can be
of order of $(10^3-10^4)$ $m^{-1}$, where $m$ is the mass of the scalar
field, {\it i.e.} much longer than that for a configuration obeying the
Klein-Gordon equation for a free scalar field, of order of $5\:m^{-1}$.

Oscillons naturally appear during the collapse of spherically
symmetric field configurations: if a bubble is formed at rest at the time $t=0$
with, for example, a "Gaussian" shape
\begin{equation}
\label{ans}
\phi({\bf x},0)=a\:{\rm e}^{-|{\bf
x}|^2/R^2},
\end{equation}
{\it i.e.} with an amplitude $a$ at its core and initial radius $R$,
after having radiated most of its
initial energy, the bubble settles into a quite long-lived regime,
before disappearing by quickly
radiating away its remaining energy.

The conditions required for the existence of the oscillons are, apart
from having the initial energy above a plateau energy,
essentially two \cite{osc}:
{\it i)} the initial amplitude of the field
at the core needs to be above the inflection point of the potential
in order to probe the nonlinearities of the theory and {\it ii)} the
configuration must have an initial radius $R$ bounded from below. To
explain these conditions fairly analitically (conclusions
are confirmed numerically),
we can follow ref. \cite{osc} and consider
the potential
\begin{equation}
\label{pot}
V(\phi)=\frac{m^2}{2}\phi^2-\frac{\alpha_0m}{3}\phi^3+\frac{\lambda}{4}\phi^4.
\end{equation}
A solution $\phi({\bf x},t)$ to the equation of motion has energy
\begin{equation}
\label{energy}
E\left[\phi\right]=\int\:d^3{\bf x}\:\left[\frac{1}{2}\dot{\phi}^2+
\frac{1}{2}\left(\nabla\phi\right)^2+V(\phi)\right].
\end{equation}
Since during the oscillon regime the subcritical configuration is
characterized by a slowly varying radius, we can
model the oscillon by writing
\begin{equation}
\label{model}
\phi({\bf x},t)=a(t)\:{\rm e}^{-|{\bf x}|^2/R_*^2},
\end{equation}
where the radius $R_*$ is kept fixed (a good approximation
supported by numerical analysis).

The equation of
motion for $a(t)$ can be linearized writing $a(t)=\bar{a}(t)+\delta
a(t)$, so that the fluctuation $\delta a(t)$ satisfies the linearized
equation
\begin{eqnarray}
\label{lin}
\ddot{\delta}a&=&-\om^2(\bar{a},R_*)\delta a,\nonumber\\
\om^2(\bar{a},R_*)&=&\frac{3\sqrt{2}}{4}\:\lambda\bar{a}^2-\frac{4\sqrt{6}}{
9}\:\alpha_0 m \bar{a}+m^2\left(1+\frac{3}{m^2 R_*^2}\right).
\end{eqnarray}
For $\om^2(\bar{a},R_*)<0$ fluctuations about $\bar{a}$ are unstable,
driving the amplitude away from its vacuum value and thus avoiding
the rapid shrinking of the initial configuration. These are the
fluctuations responsible for the appearance of the oscillon and its
relative long lifetime \cite{osc}.

For fixed
$\alpha_0$, $\om^2(\bar{a},R_*)$ does have a minimum for $\bar{a}_{\rm
min}\simeq 0.51\:(\alpha_0 m/\lambda)$, hence oscillons are possible
only for
\begin{equation}
\label{limit}
R_*>R_{\rm
min}\simeq\sqrt{3/\left(0.28\alpha_0^2-\lambda\right)}\:\lambda^{1/2}\: m.
\end{equation}
For $R_*>R_{\rm min}$, $\om^2(\bar{a},R_*)$ will be negative
for amplitudes
\begin{equation}
\label{limit}
\bar{a}_{-}<\bar{a}<\bar{a}_{+},
\end{equation}
where
\begin{equation}
\label{values}
\bar{a}_{\pm}=\frac{8\sqrt{3}}{27}\frac{\alpha_0 m}{\lambda}
\pm\frac{\sqrt{2}}{3}\left[\frac{96}{81}\frac{\alpha_0^2 m^2}{
\lambda^2}-3\sqrt{2}\left(1+\frac{3}{m^2
R^2}\right)\frac{m^2}{\lambda}\right]^{1/2}.
\end{equation}
In the limit of very large $R_*$, $R_*\gg 1/m$, $\bar{a}_{-}$ becomes
independent from $R_*$. To give a numerical example, in the
degenerate case $\alpha_0=(3/\sqrt{2})\:\lambda^{1/2}$
\begin{equation}
\label{nu}
\bar{a}_{{\rm inf}}\simeq 0.3\:\frac{m}{\sqrt{\lambda}}<\bar{a}_{-}
\simeq 0.6\: \frac{m}{\sqrt{\lambda}}<\bar{a}_{{\rm max}}\simeq
0.7\:\frac{m}{\sqrt{\lambda}},
\end{equation}
where we have indicated with $\bar{a}_{{\rm inf}}$ and $\bar{a}_{{\rm
max}}$ the inflection point closest to $\bar{a}=0$ and the maximum
of the potential $V(\bar{a})$, respectively.

It is then clear why oscillons can  form only if their
 initial amplitude at the
core is above the inflection point $a_{{\rm inf}}$ and why their
initial radius $R$ cannot be too small in order to feel
the nonlinearities of the potential.

One of the motivations for studying the evolution of unstable
spherically-symmetric configurations comes from the original papers
analyzing the
role subcritical bubbles may play in the dynamics of weak first order
phase transitions \cite{sub}. Considering models with double-well
potentials in which the system starts localized
on one minimum, for sufficiently weak transitions correlation-volumes
bubbles of the other phase could be thermally nucleated, giving rise to
an effective phase mixing between the two available phases before the
reaching from above of the tunneling temperature at which
critical bubbles are expected to be nucleated. This could have
dramatic consequences for any electroweak baryogenesis mechanisms
\cite{ckn}.

Although the presence of thermal fluctuations in any hot system is
undisputed, their role in the dynamics of weakly first order phase
transitions is still under debate \cite{contra}. However, it is clear
that, {\it if} thermally nucleated, long-lived oscillons
could appreciably affect
the dynamics of a weak first order phase transition at the weak scale.
Although their lifetime is small in comparison with the expansion
time-scale for temperatures $T\sim 100$ GeV,
if oscillons are produced in large enough numbers, their
presence will substantially increase the equilibrium number-density of
subcritical bubbles of the broken phase. This could effectively make the
transition weaker than what predicted from the effective potential.
Also, instabilities on the expanding critical bubble walls could
be generated by collisions with oscillons, implying that the usual
assumption of spherical evolution of the walls may be incorrect.

The aim of this Letter is to investigate whether oscillons
can  be really present
at the onset of a first order  electroweak phase
transition, {\it i.e} if subcritical bubbles with initial amplitude
at their core and initial radius $R$ satisfying the above conditions
{\it i)} and {\it ii)} can be thermally nucleated and
affect the usual picture of the phase transition dynamics.

To answer this
question we will treat both the initial amplitude and the size of subcritical
bubble as statistical degrees of freedom along the same lines of what
done by Enqvist {\it et al.} in ref. \cite{contra}.

First order phase transition and bubble dynamics in the Standard Model
have lately been studied in much detail, and it has become increasingly clear
\cite{clear} that for Higgs masses
considerably heavier than 60 GeV, the electroweak phase transition
is only of weakly first order. For Higgs mass
$M_H > 100$ GeV the calculations, both perturbative and lattice ones, confront
technical problems and it is conceivable that for such large
Higgs masses the electroweak phase transition is close to a second order
and does not proceed by critical bubble formation.
Therefore, in this Letter we use a
phenomenological Higgs potential for the order parameter $\phi$ suitable
for a simple description of a first order phase transition:
\be{potential}
V(\phi,T) = \frac 12 m^2(T)\phi^2 - \frac 13 \alpha T \phi^3 + \frac 14
\lambda\phi^4.
\ee
The properties of the oscillons for the potential (\ref{potential})
can be easily derived from the analysis made for the
potential (\ref{pot}) with the substitution $\alpha_0 m\rightarrow
\alpha T$. Namely, the oscillon stage can be obtained only if
subcritical configurations have initial amplitude greater  than the
inflection point $\phi_{{\rm inf}}$ and sufficiently large size. We also
expect that, when increasing $\lambda$, the minimum necessary value for $R$
increases, whereas the smallest available value of the amplitude
at the core $\bar{\phi}_{-}$ decreases \cite{osc}.

When discussing oscillons one has to be sure that initial
configurations, which eventually will give rise to an oscillon, are not
critical bubbles. Indeed, for the potential (\ref{potential}) and at
tunneling temperature $T_f$, critical bubbles become  solutions
of the equation of motion: if they are nucleated with an initial  radius
$R_c$ (or larger) they can grow converting the metastable phase $\phi=0$
into the stable phase with lower energy.

Most of the dynamical properties of the electroweak
phase transition associated with the potential \eq{potential},
such as the smallness of the latent heat, the bubble
nucleation rate and the size of critical bubbles, have been discussed
extensively in \cite{kari}. For the purposes of the present paper it suffices
to
recall only some of the results.

Assuming that there is
only little supercooling, as seems to be the case for the electroweak
phase transition, the bounce action can be written as
\be{bounce}
S/T = {\alpha\over \lambda^{3/2}} {2^{9/2}\pi\over 3^5}
{\bar\lambda^{3/2}\over
(\bar\lambda - 1)^2}~,
\ee
where $\bar\lambda (T)=9\lambda m^2(T)/(2\alpha^2T^2)$.
The cosmological transition temperature is determined
from the relation that the Hubble rate equals the transition rate $\propto
{\rm e}^{-S/T}$,
yielding $S/T_f \simeq {\rm ln} (M_{{\rm Pl}}^4/T_f^4) \simeq 150$,
where $T_f$ is the transition temperature. Thus we obtain from \eq{bounce}
\be{lambdabar}
\bar \lambda(T_f) \simeq 1 - 0.0442{\alpha^{1/2}\over\lambda^{3/4}}\equiv
1-\delta.
\ee
On the other hand, small supercooling implies that $1-\bar\lambda=\delta\ll 1$,
i.e. $\alpha \ll 500 \lambda^{3/2}$. Solving for $\bar\lambda$ in \eq{bounce}
yields
the transition temperature $T_f$. One finds
\be{mass}
m^2(T_f) = {2 \alpha^2\over 9\lambda}\:\bar\lambda(T_f) \: T_f^2~.
\ee
The extrema of the potential are given by
\be{extrema}
\phi_{\rm min,max}(T) =
{\alpha T\over 2\lambda}\left(1 \pm\sqrt{1 - 8\bar\lambda/9}\right).
\ee
Expanding the potential at the broken minimum $\phi_{{\rm min}}(T)$ we find
\be{epsilon}
-\epsilon\equiv V(\phi_{{\rm min}},T_f)={1\over 6}m^2(T_f)\phi_{{\rm
min}}^2-
{1\over 12}\lambda\phi_{{\rm min}}^4=-0.00218\:{\alpha^{9/2}\over
\lambda^{15/4}}
T_f^4+{\cal O}\left(\delta^2\right).
\ee
The height of the barrier is situated at $\phi_{{\rm max}}\simeq
\phi_{{\rm min}}/2$ with $V\left(\phi_{{\rm min}},T_c\right)\equiv V_{{\rm
max}}=\alpha^4 T_c^4/(144\:\lambda^3)$, where $T_c$ is the temperature
at which $V(0)=V(\phi_{{\rm min}})$, given by the condition $m(T_c)^2=(2\:
\alpha^2\: T_c^2/9\: \lambda)$. As $T_c\simeq T_f$ we may conclude that
the thin wall approximation is valid if $-\epsilon/V_{{\rm max}}=
0.314\:\alpha^{1/2}/\lambda^{3/4}\ll1$, or
$\alpha\ll 10\lambda^{3/2}$. Thus  the small supercooling
limit is clearly satisfied if the thin wall approximation is valid.

The closest inflection point to $\phi=0$ at $T\simeq T_c$ is given by
\begin{equation}
\phi_{{\rm inf}}\simeq 0.42\: \phi_{{\rm max}}.
\end{equation}

To get the size of the critical bubble we still need the surface
tension. One easily finds
\be{surface}
\sigma=\int_{0}^{\infty}\:d\phi\sqrt{2\:V(T_c)}={2\:\sqrt{2}\:\alpha^3\over
91\:\lambda^{5/2}}\:T_c^3.
\ee
We define the critical bubble radius by extremizing the
bounce action. The result is
\be{Rc}
R_c = 13.4\, {\lambda^{3/4}\over\alpha^{1/2}m(T_f)}.
\ee
Therefore $R_c$ is much
larger than the correlation length $\xi(T_f) = 1/m(T_f)$ at the transition
temperature, as it should.

Let us first make the general observation that
it is the actual transition temperature $T_f$ rather than the critical
temperature $T_c$
which is relevant for the study of oscillons.
This is true in the sense that if oscillons are not important
at $T_f$, they most certainly will not be so at $T_c$. As we shall show,
it actually turns out that oscillons are not present even
at  $T_f$. This justifies, in retrospect, our choice $T=T_f$ for performing
the calculations.

In the case of a
weak first order phase transition the critical bubble is typically well
described by a thin wall approximation, where the  configuration
has a flat 'highland' (with $\phi$ determined
by the non-zero minimum of the potential) and a steep slope down to
$\phi=0$. Therefore it seems natural that also a large subcritical
bubble should resemble the critical one, {\it i.e.} when $R$ increases,
the form of the subcritical bubble should deform smoothly so that, when
$R=R_c$, the bubble is a critical one.

Motivated by this observation, let us  define a
subcritical bubble as a functional of both the amplitude $a$
and the radius $R$. For this purpose one has first to
study the behaviour of the
potential as a function of the amplitude. At $T_f$ there is a interval $\phi
\in [a_-,\ a_+]$ where $V(\phi ) \le 0$. If the amplitude of the bubble is
in that interval, there exists a
critical bubble-solution of the bounce action. This means that
we have a relation
$R_c = R_c(a)$ which reproduces \eq{Rc} if $a = \phi_{{\rm min}}$. Therefore
$R_c(a)$
serves as an upper limit for the initial radius $R$ of a subcritical
bubble in that region: if
$R>R_c(a)$ we exclude such a configuration since it should give rise
to a critical bubble and not, eventually, to an oscillon with finite
lifetime.

These considerations lead us to
define different Ans\" atze for various regions in the $(a, R)$-plane. When
$\phi\in [a_-,\ a_+]$, we use an Ansatz such that when
$R \rightarrow R_c(a)$, the
field configuration goes towards the thin wall form. For small $R$ we use a
simple gaussian configuration. For other values of $a$ we always take a
thin-wall like Ansatz. Thus we write for $\phi \in [a_-,\ a_+]$ and $R \leq
R_c(a)$
\be{ansatz1}
\phi(t,\ R) = a(t) \left[{R_c(a) - R\over R_c(a)}\phi_g + {R\over
R_c(a)}\phi_t\right],
\ee
where $t$ is the time coordinate\footnote{Note, however, that we need not to
specify the explicit time evolution of $a$ and $R$ when dealing with
statistical averages.} and
\bea{functions}
\phi_g(R) &=& {\rm e}^{-r^2/R^2},\\
\phi_t(R) &=& 1/({\rm e}^{m(r - R)} + 1),\\
r&=&|{\bf x}|,
\eea
Such an Ansatz reproduces the requirement that when $R\rightarrow R_c(a)$,
subcritical
bubbles should resemble  critical ones. In practise the statistical
averages depend only weakly on $a$ because the main contribution to them
comes from the region of small $a$ amd large $R$. Therefore we assume
for simplicity that criticality depends weakly on $a$ and take $R_c(a)=
R_c$ to be a constant whenever possible.

For $\phi \not\in [a_-,\ a_+]$ we assume that no gaussian component is
present and write simply
\be{ansatz2}
\phi(t,\ R) = a(t)\phi_t(R).
\ee
However, the statistical averages are expected to be quite insensitive of
the precise form of the configuration.

These Ans\" atze can be plugged into the action
\be{action}
S[a,\ R] = \int d^4x\, \left[\frac 12 (\partial\phi)^2 - V(\phi)\right]
\ee
from which the Lagrangian in terms of the dynamical variables $a$ and $R$ can
be extracted. In the practical calculation we have, whenever possible,
approximated $\phi_t$ by the step function.
After that is a simple matter to calculate the effective
Hamiltonian function $H_{{\rm eff}}$ of the
dynamical variables $a$ and $R$.

Once we have the Hamiltonian, we may calculate the
statistical average of a dynamical variable of the type $F(a, R)$ simply by
\be{generalave}
\langle F(a, R)\rangle = {\int dp_R\, dp_a\, da\, dR \,F(a, R)
{\rm e}^{-\beta H_{{\rm eff}}}
\over \int dp_R\, dp_a\, da\, dR\, {\rm e}^{-\beta H_{{\rm eff}}}}.
\ee
However, because the effective Lagrangian is of the form
\be{lag}
{\cal L}_{{\rm eff}} = \frac 12 \pmatrix{\dot a & \dot R\cr}
K \pmatrix{\dot a \cr \dot R\cr} - {\cal V},
\ee
where $K=K(a,R)$ is a symmetric matrix,
after the momentum integration the average
can be cast into the form
\be{ave}
\langle F(a, R)\rangle = {\int da\, dR \,F(a, R)
\sqrt{\det K} {\rm e}^{-\beta {\cal V}}\over \int da\, dR\, \sqrt{\det K}
{\rm e}^{-\beta
{\cal V}}}.
\ee
The matrix
\be{K}
K = 4\pi \pmatrix{K_{11}& K_{12}\cr
           K_{21}& K_{22}\cr}
\ee
and the pseudopotential ${\cal V}$ are given separately for the two
regions. For $\phi \in [a_-,\ a_+]$ we obtain
\bea{Kinreg.1}
R_c^2\, K_{11} &=& \Delta^2 R^3 A^2_2 + 2 \Delta R^4 B_2^1 + \frac 13 R^5
\nonumber\\
R_c^2 \,K_{12} &=&  2 a \Delta^2 R^2 A_4^2 - a \Delta R^3 A_2^2 + \frac 13 a
R^4 +
a R^5 m I(mR) \nonumber\\
& & + a \Delta R^3 B_2^1 - a R^4 B_2^1 + 2 a \Delta R^3 B_4^1
+a \Delta R^4 m J_2(mR)\nonumber \\
R_c^2\, K_{22} &=& 4 a^2 \Delta^2 R A_6^2 + a^2 R^5 m^2 I(2mR) + 4 a^2 \Delta
R^3 m
J_4(mR) \nonumber\\
& & + a^2 R^3 A_2^2 - 2 a^2 R^3 B_2^1
+ \frac 13 a^2 R^3 - a^2\Delta R^2A_4^2 \nonumber\\
& & - 2a^2 R^4mJ_2(mR) + 4a^2
\Delta R^2 B_4^1 + 2a^2 R^4 m I(mR)
\eea
and
\bea{Linreg.1}
{R_c(a)^2\over 4\pi}{\cal V} &=&  2 a^2\Delta^2 R A_4^2 + 2 a^2 \Delta R^3 m
J_3(mR) + \frac 12a^2 R^5 m^2 I(2mR)\nonumber\\
 & & + \frac 12 m^2 a^2 \Delta^2 R^3 A_2^2 +
m^2 a^2 \Delta R^4 B_2^1 + \frac 16 m^2a^2 R^5 -
\frac 13 \alpha T {a^3\over R_c(a)}\Delta^3R^3A_2^3\nonumber\\
& & - \alpha T {a^3\over R_c(a)}\Delta^2 R^4 B_2^2 - \alpha T {a^3\over R_c(a)}
\Delta R^5 B_2^1 - \frac 19 \alpha T {a^3\over R_c(a)}  R^6\nonumber\\
& & + \frac 14 \lambda {a^4\over R_c(a)^2} \Delta^4 R^3 A_2^4
 + \lambda {a^4\over R_c(a)^2} \Delta^3 R^4 B_2^3
+ \frac 32 \lambda {a^4\over R_c(a)^2} \Delta^2 R^5 B_2^2\nonumber \\
& & + \lambda {a^4\over R_c(a)^2} \Delta  R^6 B_2^1
+ \frac 1{12} \lambda {a^4\over R_c(a)^2}  R^7.
\eea
Note that in \eq{Linreg.1} the $a$ -dependence of $R_c$ has to be used
explicitly
because the critical behaviour is determined from it.
For the region where $\phi \not\in [a_-,\ a_+]$ the corresponding functions
are given by
\bea{Kinreg.2}
K_{11} &=& \frac 13 R^3 \nonumber\\
K_{12} &=& a R^3 m I(mR)\nonumber \\
K_{22} &=&  a^2 R^3 m^2 I(2mR)
\eea
and
\be{Linreg.2}
{1\over 4\pi}{\cal V} =  \frac 12 a^2 R^3 m^2 I(2mR) + \frac 16 m^2a^2 R^3 -
 \frac 19 \alpha T a^3  R^3 + \frac 1{12} \lambda a^4  R^3.
\ee
A number of shorthand notations have been introduced in the previous equations:
\bea{notations}
\Delta &=& R_c(a) - R\\
A_n^k &=& \int_0^\infty du\, u^n e^{- k u^2} = {\Gamma({n+1\over 2})
\over 2 k^{n+1\over 2}}\\
B_n^k &=& \int_0^1 du\, u^n e^{- k u^2} \\
I(x) &=& \int_0^1 du\, u^2 e^{x(u -1)} = \frac 1x - \frac 2 {x^2} +
\frac 2{x^3} - \frac 2{x^3} e^{-x}\\
J_n(x) &=& \int_0^1 du\, u^n e^{- u^2 + x(u-1)}.
\eea

The range of integration for $R$ posses un upper limit given by
thermalization.
Motivated by the fact that thermal fluctuations can generate configurations
with spatial size comparable to the critical bubble radius, which may
affect the dynamics of a first order phase transition,
the authors of ref.
\cite{ElmforsEV} have estimated the lifetime of fluctuations
of an on-shell
Higgs field with zero  momentum $\left(p_0=m(T),{\bf
p}=0\right)$. This choice  reflects the fact
that critical bubbles are typically
much larger than the interparticle distance $\simeq 1/T$ in plasma.
Writing $p_0\equiv \omega-i\:\gamma/2$, one finds that the dispersion
relation is
\be{dis}
\omega^2=\left|{\bf p}\right|^2+m^2(T)+{1\over 4}\gamma^2,
\ee
where
\be{gamma}
\gamma={{\rm Im}\:\Gamma^{(2)} \over\omega},
\ee
$\Gamma^{(2)}$ being the two-point function for the Higgs field.

The imaginary part arises at one loop level, but because of kinematical
constraints, the two loop contribution is actually dominant in the
region of physical couplings.
The thermalization rate $\gamma$ for small
amplitude scalar fluctuations and large spatial size, $
R\sim\left|{\bf p}\right|^{-1}\gg \gamma^{-1}$, is estimated \cite{ElmforsEV}
to be of the order $\gamma
\simeq 10^{-2}\: T$ near the critical temperature, {\it i.e.}
much larger than the typical first order transition time. This means that all
small amplitude fluctuations with size larger than
\be{maxR}
R_{{\rm max}}={\cal O}(1/\gamma)
\ee will effectively
be absent from the mixture of subcritical bubbles and must
not counted in the thermal averages. In
practise, the limit \eq{maxR} is of the order of few times $R_c$, depending on
the actual value of $\gamma $. Even if it is not precisely known, its
inclusion in the calculations is important. Without it all statistical
averages would be dominated by infinite, infinitesimally small fluctuations.
Technically this can be seen from the \eq{ave}, where the integrals
diverge in the limit $a\goto 0,\ R\goto\infty$. It is important to note that
the divergence is not a problem of our Ansatz but merely a more general
phenomenon, which seems to be related to the general infra-red instability
problems emerging in the calculations of the effective action.

We have computed the average initial
radius and the amplitude of fluctuations at
$T=T_f$ from Eq. (\ref{ave}) numerically using a cut-off $R_{{\rm
max}}\simeq 3.3\: R_c$ (we have checked numerically that
results do not change significantly for
different choices of $R_{{\rm max}}$).

We have taken  $\alpha=0.048$
and varied $\lambda$ between $4\times 10^{-2}$ and $10^{-1}$.
For larger values of $\lambda$ the first
order electroweak phase transition is close to a second order and does
not proceed by critical bubble formation.

For instance, when our phenomenological potential
\eq{potential} is fitted to the two loop result for the effective
potential calculated in \cite{lattice} for the Higgs mass $M_H=70$ GeV,
this yields $\lambda\simeq 0.061$. One can
readily verify that the thin wall approximation is valid in the chosen
range for $\lambda$.

In Fig. 1 we show the ratio between $\langle
a^2\rangle^{1/2}$ and the inflection point $\phi_{{\rm inf}}(T_f)$
as a function of $\lambda$. This ratio is always of order of
$0.5$. We have also computed numerically the ratio
$\langle a\rangle/\phi_{{\rm inf}}(T_f)$ which turns out to lie
in the range (0.13--0.17) in the given range for $\lambda$.
We recall that the oscillon stage can be present only if subcritical
bubbles are thermally nucleated with initial amplitude above the
inflection point \cite{osc}.

In Fig. 2 we show $\langle R^2\rangle^{1/2}$ in units of $R_c$ and
we have numerically computed $\langle R\rangle$ to be in the range
(1.74--1.58) $R_c$. Oscillons can be formed
only if $R>R_{{\rm min}}\simeq
3/m(T_f)\simeq (0.55-0.27)\:R_c$ \cite{osc}.

Note that, when $\lambda$ increases, the average amplitude at the core
also increses, whereas the smallest available amplitude
$\bar{\phi}_{-}$ to settle into an oscillon
stage decreaes \cite{osc}. However, in spite of this tendency, the
avarage amplitude is always smaller than $\bar{\phi}_{-}$ for any chosen
value of $\lambda$.

Thus, even if our
results seem to indicate that condition {\it ii)} is satisfied, {\it
i.e.} subcritical bubbles can be thermally nucleated with sufficiently large
average initial radius $R$ to give rise to the oscillon regime,
nevertheless initial average amplitudes at the core do not satisfy
condition {\it i)} since they always result to be smaller than the inflection
point $\phi_{{\rm inf}}(T_f)$.

Thermal fluctuations are certainly present at the onset of the
electroweak phase transition, but the most probable subcritical configuration
generated around the critical temperature, even if with sufficiently
large size, is charaterized by an amplitude too small to begin the
oscillon stage. From this result we can infer that the dynamics of a weak
electroweak first
order phase transition is not affected by the presence of
long-lived oscillons, suggesting that  the electroweak baryogenesis scenarios
are still viable to explain the generation of the baryon asymmetry
in the early Universe.  However, we feel that an important issue
deserves further study:
relaxation time-scales of subcritical configurations depend on the nature of
the
stochastic force and the strength of dissipation provided by the
surrounding thermal bath and their complete
knowledge is needed to decide if degrees of freedom
are in equilibrium or not inside the subcritical bubble.

One may realize that this is a
crucial question by reminding that the effective
potential (\ref{potential}), used to describe the free
energy associated to the fluctuations,
is usually obtained integrating out fermionic
and the bosonic degrees of freedom of the
theory.

In
performing such a calculation, it is commonly assumed that fermions and
bosons do have {\it equilibrium} distributions with a $\phi({\bf x},t)$
background
dependent mass. This is true only if their interaction times with the
background $\phi({\bf x},t)$ are much smaller than the typical lifetime of the
subcritical bubble.

Since this condition is not always satisfied, a full
non-equilibrium approach is needed. The latter, however, seems to
confirm, or even strengthen,  the results of this paper
about oscillons \cite{iiro}.

\newpage

\newpage
\begin{flushleft}
{\large\bf Figure captions}
\end{flushleft}
\begin{flushleft}
{\bf Figure 1} The plot of the ratio $\langle
a^2\rangle^{1/2}/\phi_{{\rm inf}}(T_f)$ as a  function of $\lambda$ and
for $\alpha=0.048$.
\end{flushleft}
\begin{flushleft}
{\bf Figure 2} The plot of $\langle
R^2\rangle^{1/2}$ in units uf $R_c$ as a  function of $\lambda$ and
for $\alpha=0.048$.
\end{flushleft}
\end{document}